\documentclass[pdflatex,sn-mathphys-ay]{sn-jnl}

\usepackage[T1]{fontenc}%
\usepackage{graphicx}%
\usepackage{amsmath,amssymb,amsfonts}%
\usepackage[title]{appendix}%
\usepackage{array}%
\usepackage{xcolor}%
\usepackage{textcomp}%
\usepackage{booktabs}%

\newcommand{\Kbar}{\bar{K}}
\newcommand{\dd}{\mathrm{d}}
\newcommand{\asec}{$''$\,yr$^{-1}$}
\newcommand{\whf}{\texttt{WHFast512}}
\newcommand{\rebound}{\texttt{REBOUND}}

\begin{document}

\title[Mean-element parareal for the Solar System]{Mean-element Parareal: Parallel-in-Time Integration of the Solar System}

\author*[1]{\fnm{Michael} \sur{Hippke}}\email{michael@hippke.org}

\affil*[1]{\orgname{Sonneberg Observatory}, \orgaddress{\street{Sternwartestr.~32}, \city{Sonneberg}, \postcode{96515}, \country{Germany}}}

\abstract{\unboldmath Long-term integrations of planetary systems have long been considered inherently sequential: each time step must start from the \emph{exact} state at the end of the previous one, because any error in that state, however small, is amplified by orbital shear and chaos and alters the trajectory. Because a single step of a modern Wisdom--Holman integrator takes only $\sim$100\,ns---too short to split across cores---additional cores have previously only been used to run multiple simulations in parallel. We introduce mean-element parareal (MEP), a parallel-in-time scheme that accelerates a single Solar System integration (eight planets, general relativity) at the same accuracy as the serial integrator. Combined with a latency-optimized C kernel that integrates 5\,Gyr on one core in 5.8\,h, MEP completes a 5\,Gyr integration in 52\,min on 64 cores (a parallel speed-up of $6.7\times$, and $10.4\times$ faster than the previous 9.0\,h single-core record) or in 1.8\,h on 16 cores ($3.3\times$ parallel speed-up, $5.0\times$ faster than the previous record). MEP succeeds because its coarse propagator uses a numerical first-order mean-element theory to remove the short-period oscillations from the semi-major axes, preventing Keplerian shear from amplifying them into growing phase errors. Standard parareal with a coarse Wisdom--Holman integrator fails for planetary orbits because no inexpensive coarse integrator can resolve these oscillations. To our knowledge, this is the first measured multicore speed-up of a single long-term planetary integration and opens a pathway towards larger-scale parallelization. MEP is freely available in Parachron, a modified version of \rebound{}, at \url{https://github.com/hippke/parachron}.}


\maketitle

%% ======================================================================
\section{Introduction} \label{sec:intro}
%% ======================================================================

Studying the long-term dynamical evolution of the Solar System and of exoplanetary systems requires integrating the gravitational $N$-body problem over billions of orbits \citep[e.g.,][]{SussmanWisdom1988,Laskar1989,SussmanWisdom1992,Laskar2008,BatyginLaughlin2008,LaskarGastineau2009,Zeebe2015,Abbot2023,BrownRein2023,ReinBrownKanda2025}. For these integrations, mixed-variable symplectic maps of the Wisdom--Holman (WH) type \citep{WisdomHolman1991,Kinoshita1991} have become the standard tool, and successive implementations have steadily reduced their computational cost: from \texttt{WHFast} \citep{ReinTamayo2015} in the \rebound{} package \citep{ReinLiu2012} to the AVX-512 vectorized \whf{} \citep{Javaheri2023} and its hand-written assembly successor, which integrates the eight planets of the Solar System with general relativistic (GR) corrections for 5\,Gyr in 9.0\,h on a single core \citep{DagliRein2026}. As \citet{DagliRein2026} argue, this code already approaches the hardware limit of a single core on present-day CPUs.

With single-core performance near its ceiling, further speed-ups must come from parallel hardware, yet conventional parallelism has not accelerated a single planetary integration. Because every time step must start from the exact end state of the previous one, the integration is inherently sequential; moreover, a single WH step takes only $\sim$100\,ns, comparable to the core-to-core latency of a CPU, leaving too little work within a step to distribute across cores. Distributing force evaluations across particles pays for large $N$ \citep{GrimmStadel2014,Grimm2022,ZhangGladman2022,LauLee2023}, but for few bodies a 16-core CPU outperforms a GPU by at least a factor of ten \citep{GrimmStadel2026}, and vectorization within a step is already what makes \whf{} fast \citep{Javaheri2023}. Parallel computers have therefore been confined to running independent trajectories in ensembles \citep[e.g.,][]{LaskarGastineau2009,BrownRein2022,Abbot2023} or parameter surveys \citep[e.g.,][]{Tamayo2016ml,Obertas2017,Tamayo2020}, while the wall-clock time of any individual trajectory remains fixed by single-core speed. Averaged (secular) equations of motion run orders of magnitude faster than direct integration \citep{Laskar1990,Laskar2008,MogaveroLaskar2021,Hoang2022}, but only by approximating the dynamics. Many questions instead demand exact, ultra-long single trajectories: determining the dynamical half-life of Mercury over 100\,Gyr \citep{Hoang2022}, constructing orbital solutions for astrochronology \citep{Laskar2004,Zeebe2023}, probing slow chaotic diffusion \citep{Mogavero2023}, and establishing the numerical convergence of Solar System ensembles \citep{HernandezZeebeHadden2022,ReinBrownKanda2025,ReinDey2026}.

\begin{figure}
\centering
\includegraphics[width=0.7\textwidth]{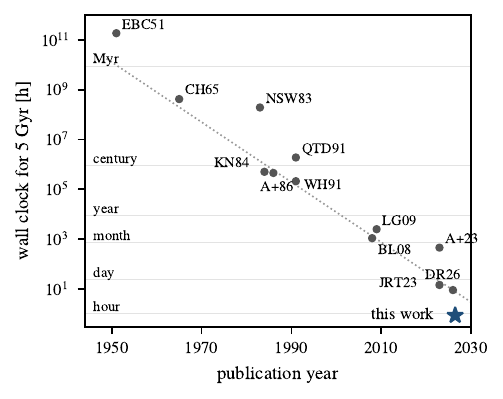}
\caption{Wall-clock time to integrate the eight planets of the Solar System for 5\,Gyr, by publication year, normalized as in \citet[][their figure~1]{DagliRein2026}: $t_{\rm wall}\,(5\,{\rm Gyr}/t_{\rm sim})\,(P_{\min}/88\,{\rm d})\,(8/N)^2$. Grey points: published integrations, compiled by S.~Hadden (\url{https://shadden.github.io/nbody_history/}) and re-derived with this normalization (approximate); JRT23 is inferred from the statement of \citet{DagliRein2026} that their version of \whf{} is 60\% faster than that of \citet{Javaheri2023}. Dotted line: halving every 2.5\,yr. This work (blue star): MEP at the same accuracy as the serial integrator, taking 52\,min (0.9\,h) on 64 cores ($10.4\times$ faster than the 9.0\,h record of \citealt{DagliRein2026}, DR26) and 1.8\,h on 16 cores ($5.0\times$ faster). Serial \whf{} with Parachron's C kernel takes 5.8\,h on one core.}
\label{fig:history}
\end{figure}

In this paper, we show that a single long-term integration of the Solar System can be parallelized in time on a many-core workstation. Our method, mean-element parareal (MEP), integrates the benchmark of \citet{DagliRein2026}---the Sun and eight planets with GR corrections at a 6\,d time step---for 5\,Gyr in 52\,min on 64 cores and 1.8\,h on 16 cores, compared to 5.8\,h on a single core with our C kernel (speed-ups of 6.7 and 3.3, or $10.4\times$ and $5.0\times$ faster than the 9.0\,h record of \citealt{DagliRein2026}) at the full accuracy of the serial integrator (Figure~\ref{fig:history}). Across billion-year integrations, the parallel trajectories preserve the energy and angular-momentum conservation, the secular frequencies, and the long-term dynamical statistics of the serial trajectory.

To parallelize along the time axis, parallel-in-time (PinT) methods, of which parareal \citep{Lions2001} is the best known, divide the time span into slices, integrate many slices concurrently from approximate initial states, and iteratively correct the slice boundaries using a fast coarse propagator \citep[see the reviews of][]{Gander2015,OngSchroder2020}. Although their application to planetary orbits has a long history, no wall-clock speed-up has previously been measured. \citet{SahaStadelTremaine1997} noted that ``simulations on serial machines suggest'' a speed-up of about 50 on 1000 processors; \citet{JimenezPerezLaskar2011} tested a refined scheme on model problems in a simulated parallel environment; \citet{Dai2013} reported a theoretical speed-up above 60 for the outer Solar System ``provided a sufficient number of processors are available'', requiring up to 15 iterations; and \citet{GanderHairer2014} proved for the Kepler problem that rapid convergence demands a coarse solver of nearly the fine solver's accuracy. These difficulties reflect three broader obstacles documented in the PinT literature: parareal struggles with oscillatory problems because coarse phase errors stall convergence \citep{Ruprecht2018}, loses the geometric properties of the fine solver unless symmetrized or projected \citep{BalWu2008,Dai2013}, and converges only over a limited interval for chaotic systems \citep{GanderHairer2008}.

We identify the physical mechanism behind this barrier and show how to overcome it. For planetary orbits, parareal convergence is governed by how accurately the coarse propagator resolves the \emph{semi-major axes}. Through Keplerian shear, the dependence of mean motion on semi-major axis that disperses planetary rings and debris clouds along their orbits \citep[e.g.,][]{ReinTremaine2011}, any discrepancy in the semi-major axis accumulates into a secularly growing phase error. However, the osculating semi-major axes are dominated by short-period fluctuations that no computationally inexpensive coarse integrator can resolve. For oscillatory problems in other fields, the most successful PinT remedies modify the coarse model rather than its time step: they align phases \citep{ArielKimTsai2016}, formulate micro--macro models in slow variables \citep{LegollLelievreSamaey2013}, or average over the fast oscillations \citep{HautWingate2014,PeddleHautWingate2019}. Adopting this principle, we build the first PinT coarse propagator based on a perturbation theory of celestial mechanics. By transforming to \emph{mean} elements and removing the short-period terms, a low-cost coarse propagator meets the stringent semi-major-axis accuracy requirement. Perturbation theory thus steers the parallel convergence, while direct numerical integration guarantees that the converged result remains exact.

The remainder of this paper is organized as follows. Section~\ref{sec:method} presents mean-element parareal, explains the physical mechanism that governs its convergence, and describes its parallel implementation. Section~\ref{sec:accuracy} validates the physical and statistical accuracy of the parallel trajectories, and Section~\ref{sec:performance} reports the measured wall-clock speed-up and scaling. Finally, Section~\ref{sec:discussion} discusses when the method is advantageous and how it can scale to larger core counts. MEP is freely available for Python and C in Parachron, a modified version of \rebound{}.

%% ======================================================================
\section{Mean-element parareal} \label{sec:method}
%% ======================================================================

\begin{figure*}
\centering
\includegraphics[width=\textwidth]{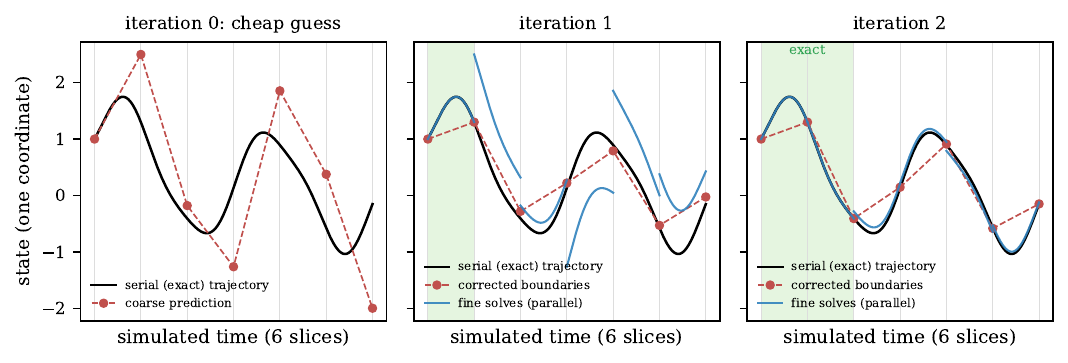}
\caption{Parareal on a toy problem with a one-dimensional state, six time slices and a coarse step of one Euler step per slice. Iteration 0: the coarse propagator predicts the slice boundaries (red), cheaply but inaccurately. Iteration $k\ge1$: all six fine solves (blue) start in parallel from the boundaries of the previous iterate, and Equation~(\ref{eq:parareal}) corrects the boundaries. Green: slices that are already exact. Two iterations suffice here to approach the serial trajectory (black) closely. In MEP the state has 48 dimensions (six orbital elements of each of eight planets), a window has 16 slices, the coarse step is the mean-element propagator of Section~\ref{sec:coarse}, and windows need 3--4 iterations.}
\label{fig:parareal}
\end{figure*}

\subsection{Fine propagator and accuracy requirements} \label{sec:contract}
Our fine propagator $F$ integrates the Sun and the eight planets in Jacobi coordinates using the Wisdom--Holman map of \whf{} \citep{Javaheri2023,DagliRein2026} with a time step of $\Delta t=6$\,d, a 17th-order symplectic corrector, and the general relativistic (GR) potential of \citet{NobiliRoxburgh1986}. Applied to a single time slice, $F$ is a standard serial integration with this map. In Parachron, we port the hand-written AVX-512 assembly kernel of \citet{DagliRein2026} to C and shorten its critical instruction dependency chain in both the Kepler solver and the Jacobi coordinate transformations (Appendix~\ref{app:asm}, Figure~\ref{fig:schedule}). Where AVX-512 is available, the compiled C kernel is faster than the assembly kernel (69 against 86\,ns per step) and runs on any processor architecture with bitwise identical results; on CPUs without AVX-512, where \rebound{} provides no \whf{}, it runs 44\% slower than with AVX-512 but still six times faster than the older \texttt{WHFast} \citep{ReinTamayo2015}. MEP works with either kernel, and all results in this paper use the C kernel. At the start and end of every slice, $F$ maps between real and symplectic-corrector coordinates ($\sim$0.06\% of the slice cost) and converts between inertial and Jacobi coordinates using recursive transformations rather than precomputed matrices, which eliminates a small energy offset per call (Appendix~\ref{app:syncbias}) and otherwise alters trajectories only at the round-off level.

Because the Solar System is chaotic with a Lyapunov time of $\sim$5\,Myr \citep{Laskar1989,SussmanWisdom1992}, two serial integrations that differ only in floating-point round-off decorrelate within $\sim$100\,Myr \citep[cf.][]{PortegiesZwartBoekholt2014}. Pointwise agreement with a single serial trajectory over Gyr timescales is therefore not a meaningful requirement. Instead, a valid parallel integration must satisfy four physical criteria: it must (i) introduce no systematic energy drift, (ii) keep energy and angular-momentum errors within the bounds of the serial integrator, (iii) reproduce the secular orbital frequencies to within the serial integrator's own time-step error, and (iv) preserve the statistical distribution of the long-term evolution, particularly Mercury's high-eccentricity excursions.

\subsection{Parareal} \label{sec:parareal}
Parareal divides a time span into slices of length $T$ and combines the accurate fine propagator $F$ with a fast, approximate coarse propagator $G$ (Figure~\ref{fig:parareal}). Starting from an initial sequential coarse prediction $U^0_{n+1}=G(U^0_n)$, iteration $k$ advances the state $U^k_n$ at the start of slice $n$ according to
\begin{equation}
U^{k+1}_{n+1} = G\!\left(U^{k+1}_n\right) + F\!\left(U^k_n\right) - G\!\left(U^k_n\right), \qquad U^k_0 = U_0.
\label{eq:parareal}
\end{equation}
Within each iteration, all fine integrations $F(U^k_n)$ are independent and run in parallel across cores; only the coarse correction sweep is sequential. After $k$ iterations, the first $k$ slice boundaries match the serial fine trajectory exactly. Once the iteration converges ($U^{k+1}_n = U^k_n$), the coarse terms $G(U^{k+1}_n)$ and $G(U^k_n)$ cancel, leaving the exact serial trajectory $U_{n+1} = F(U_n)$ as the fixed point regardless of approximations in $G$: an inaccurate coarse model costs extra iterations, not final accuracy.

With pipelining across $N$ slices (Section~\ref{sec:pipeline}), $K$ iterations, a fine slice cost $T_F$, and a coarse slice cost $T_G = r T_F$, the wall-clock speed-up is (Appendix~\ref{app:speedup})
\begin{equation}
S \approx \frac{N T_F}{N T_G + K\,(T_F+T_G)} \le \min\!\left(\frac{N}{K},\frac{1}{r}\right).
\label{eq:speedup}
\end{equation}
Achieving a speed-up above 3 on $N=16$ cores therefore requires both rapid convergence ($K\lesssim4$) and a very inexpensive coarse propagator ($r\lesssim0.03$). Because $G$ enters Equation~(\ref{eq:parareal}) only through the difference $G(U^{k+1}_n) - G(U^k_n)$, any state-independent offset in $G$ cancels out; rapid convergence instead requires $G$ to accurately capture how small perturbations at the start of a slice propagate to its end (Appendix~\ref{app:jacobian}).

\subsection{Why standard parareal fails: Keplerian shear} \label{sec:mechanism}
In planetary systems, the dominant sensitivity across a time slice is Keplerian shear. By Kepler's third law, a planet's mean motion scales with its semi-major axis as $n \propto a^{-3/2}$. A small fractional error $\delta a / a = \delta\ln a$ at the start of a slice of length $T$ shifts the mean motion by $\delta n/n = -\tfrac{3}{2}\,\delta\ln a$, which accumulates linearly in time into an orbital phase error in the mean longitude $\lambda$:
\begin{equation}
\delta\lambda(T) = -\frac{3}{2}\, n T\, \frac{\delta a}{a} = -\frac{3}{2}\, n T\, \delta\ln a .
\label{eq:shear}
\end{equation}
In MEP, each slice spans $T=360\,000$\,d $\approx986$\,yr ($6\times10^4$ fine steps), a duration long enough for each fine solve to amortize the overhead of the coarse step (Section~\ref{sec:windows}). Over a single slice, Mercury completes $\sim$4100 orbits, and the shear amplification factor $\tfrac{3}{2}nT$ reaches $3.9\times10^4$ for Mercury, $1.5\times10^4$ for Venus, $9.3\times10^3$ for Earth, and $4.9\times10^3$ for Mars. A fractional semi-major-axis error of just $10^{-8}$ is thus amplified into a phase error of $\sim$$4\times10^{-4}$\,rad within one slice. Perturbation tests of the slice map confirm that converging within $K\le3$ iterations requires the coarse propagator to track changes in $\ln a$ to an absolute accuracy of $\lesssim10^{-8}$, whereas all other orbital elements tolerate relative errors of $\sim$1\% (Appendix~\ref{app:jacobian}).

No inexpensive coarse integrator operating in standard (\emph{osculating}) coordinates can meet this $10^{-8}$ requirement. Mutual planetary perturbations cause the osculating semi-major axes to fluctuate on orbital and synodic periods with relative amplitudes of $10^{-6}$--$10^{-5}$---up to three orders of magnitude larger than the allowable error. Reproducing these short-period oscillations to $10^{-8}$ requires resolving every orbit with small time steps and accurate phases, which is as expensive as the fine integration itself.

This requirement explains why the most natural coarse propagator---the same WH integrator run with a larger time step $\Delta t_G$---cannot accelerate the integration. Resolving Mercury's 88\,d orbit restricts the coarse step to $\Delta t_G\lesssim16$\,d, which already costs at least $r=0.42$ of the fine step. At the same time, coarse sampling aliases the short-period oscillations, corrupting the semi-major-axis sensitivities by 1--5\% at $\Delta t_G=16$\,d and by $\sim$80\% at 24\,d. Across ten 986\,yr slices, a coarse WH propagator therefore needs 3--4 iterations at $\Delta t_G=8$--16\,d ($r=0.42$--0.77) and fails to converge altogether at $\Delta t_G=24$\,d ($r=0.26$) or over ten times longer slices; by Equation~(\ref{eq:speedup}), this yields $S=0.3$--0.5, a net slowdown. Multilevel schemes with a Kepler-only coarse solver fail for the same reason: in the Solar System benchmark of the \texttt{pySDC} framework \citep{Speck2019}, PFASST requires more than 15 iterations on average.

\subsection{A hybrid coarse propagator in mean elements} \label{sec:coarse}\label{sec:hybrid}
To achieve $10^{-8}$ accuracy in the semi-major axes at a small fraction of the fine solver's cost, our coarse propagator $G$ combines perturbation theory for the inner planets with direct coarse integration for the outer giants. For the \emph{inner planets} (Mercury--Mars), $G$ transforms the osculating Jacobi orbital elements into \emph{mean} elements by subtracting the short-period oscillations before evolving the slice (Appendix~\ref{app:gauss}). For each pair of planets, we evaluate their mutual gravitational accelerations on a regular two-dimensional grid of their orbital phases $(\lambda_i,\lambda_j)$ with all other orbital elements held fixed, convert these accelerations into element rates via Gauss' equations, and apply a two-dimensional fast Fourier transform (FFT). The zero-frequency Fourier mode gives the slow secular drift rate, while integrating the non-zero modes analytically over time yields the short-period displacements to subtract from the osculating elements. This procedure constitutes a numerical first-order Lie-series transformation \citep{Brouwer1959,Deprit1969}: because it evaluates perturbations on phase grids rather than expanding them in power series of eccentricity and inclination \citep{MurrayDermott1999}, it remains exact in $e$ and $i$ at first order in the planetary masses, which is essential for Mercury ($e\approx0.2$). Removing these short-period terms reduces the scatter in $\ln a$ for the inner planets by factors of 300--2300---from $2\times10^{-6}$ (Mercury) and $3\times10^{-5}$ (Mars) in osculating elements down to $10^{-9}$--$3\times10^{-8}$---well below the threshold required for parareal convergence.

For the \emph{giant planets} (Jupiter--Neptune), a first-order theory is insufficient because near-resonances---specifically the Jupiter--Saturn 5:2 great inequality (period $\approx880$\,yr) and the Uranus--Neptune 2:1 near-commensurability ($\approx4200$\,yr)---introduce small frequency divisors that produce large higher-order oscillations. However, because Jupiter's orbit is 50 times longer than Mercury's, the four giants can be integrated directly by the WH solver at a large time step of 360\,d (with the inner planets' masses added to the Sun) at negligible cost. Over one slice, $G$ transforms the inner planets to mean elements at the start, drifts them with their mean motions $n(\bar{a})$ and the secular forcing from the evolving giants, and transforms them back to osculating elements at the end.

Four design choices keep this coarse step both fast and unbiased. First, the Fourier tables are rebuilt from the current state at every coarse call because the short-period amplitudes depend on the eccentricities and inclinations; freezing the tables across iterations requires 6 iterations instead of 3 to reach the round-off floor. Second, the secular drift uses a symmetric trapezoidal (Heun) step rather than an Euler step, preventing one-sided coarse errors from biasing the secular frequencies. Third, the periodic transformation is applied only to $\ln a$ and $\lambda$, since the remaining orbital elements do not experience Keplerian shear, using $32\times32$ phase grids for the six inner--inner pairs and $16\times16$ grids for the eight pairs coupling an inner planet to Jupiter or Saturn (short-period perturbations from Uranus and Neptune on the inner planets' semi-major axes scale as $(m_j/M_\odot)(a_i/a_j)^3\lesssim10^{-10}$, two orders of magnitude below the $10^{-8}$ threshold, while their secular forcing is included through the giants' integration). The close Venus--Earth pair (semi-major-axis ratio 0.72), whose mutual perturbations contain high harmonics of the synodic angle $\lambda_{\rm V}-\lambda_{\rm E}$ but only low harmonics in the other direction, is instead tabulated on a $16\times64$ grid of $(\lambda_{\rm V},\lambda_{\rm V}-\lambda_{\rm E})$, which resolves synodic harmonics up to order 32 at the cost of a $32\times32$ grid (Appendix~\ref{app:gauss}). Fourth, the small frequency divisors of the transformation are evaluated with the mean motions of the mean rather than the osculating semi-major axes. With these choices, one coarse slice takes about 0.09\,ms against 4--5\,ms for a fine slice ($r\approx0.02$), and 16-slice windows converge to the round-off floor of $F$ in 3--4 iterations, each of which reduces the largest boundary update by one to two orders of magnitude.

\subsection{Boundary corrections and conservation laws} \label{sec:projection}
The parareal update in Equation~(\ref{eq:parareal}) is evaluated in osculating Jacobi orbital elements, with mean longitudes wrapped to $(-\pi,\pi]$. Applying the correction in orbital elements rather than Cartesian coordinates is essential: because coarse errors are predominantly phase errors along the orbit, adding Cartesian position and velocity vectors across a phase difference of 0.1--1\,rad cuts across the chord of the ellipse, producing unphysical semi-major axes and causing the iteration to diverge.

Like other parareal schemes for Hamiltonian systems \citep{Dai2013}, the element update alone does not strictly conserve total energy $E$ and angular momentum $\mathbf{L}$ across slice boundaries. We therefore project each corrected boundary state onto target values of $E$ and $\mathbf{L}$ defined by the start-of-slice value plus the increment measured during the most recent fine integration of that slice. This projection is accomplished by applying a tiny uniform rescaling and an infinitesimal rigid rotation to the planets' barycentric velocities (Appendix~\ref{app:jacobian}). Unlike manifold projections applied to non-symplectic integrators, here the underlying fine solver $F$ is symplectic: the projection only removes the small nonlinear mismatch ($\Delta E/E\sim10^{-14}$ at the accepted iterate) introduced by adding osculating elements in Equation~(\ref{eq:parareal}), so each slice carries only the conservation error of a single fine solve and the converged parareal fixed point remains unchanged.

\subsection{Windows, pipelining, and stopping criterion} \label{sec:algorithm}\label{sec:windows}\label{sec:pipeline}\label{sec:stopping}\label{sec:parcoarse}
Because the parareal error grows binomially with the number of slices from the last exact boundary, long intervals require more iterations to converge (3 iterations for 10 slices versus 4 for 20 or 40). We therefore integrate the trajectory in consecutive \emph{windows} of $N$ slices, with $N$ equal to the number of cores and each window starting from the converged end state of the previous one. The slice length $T=360\,000$\,d $\approx986$\,yr ($6\times10^4$ fine steps) balances the coarse error, which grows with slice length as $T^3$, against the relative overhead of the coarse step, which rises for shorter slices (Appendix~\ref{app:speedup}).

Within each window, one thread is pinned to one core and owns one slice. As soon as thread $n$ receives its updated boundary state $U^k_n$, it evaluates the coarse step $G(U^k_n)$, forms $U^k_{n+1}$ via Equation~(\ref{eq:parareal}), passes it to thread $n+1$, and immediately begins the fine solve $F(U^k_n)$ needed for iteration $k+1$. Coarse sweeps of successive iterations thus propagate through the window like a wave while the fine integrations of consecutive iterations overlap \citep[][Appendix~\ref{app:speedup}]{Minion2010}. Because the coarse sweep lies on the serial critical path, each coarse call is further accelerated by offloading its independent Fourier-grid evaluations to helper threads running on cores whose slices have already converged. All reductions are performed in a fixed order, making the parallel trajectory bitwise reproducible regardless of thread timing.

A window terminates at the first iteration $k\ge K_{\min}$ whose largest planet-position update across all slice boundaries, $u_k=\max_n|\mathbf{x}(U^k_n)-\mathbf{x}(U^{k-1}_n)|$, is at most $\tau$, and at the latest at $K_{\max}=6$. Because $u_k$ measures the error of iterate $k-1$ and each iteration contracts the error by a factor of $10$--$100$, the accepted iterate $k$ is one to two orders of magnitude more accurate than $\tau$. We use $K_{\min}=3$ and $\tau=2\times10^{-6}$\,AU. Because convergence is known only when the last slice completes its update, threads speculatively start their next fine solve and abort it once the window terminates. Once a window converges, all $N$ slice boundaries and any intra-slice outputs from the final fine pass are available in chronological order, and Parachron supports saving and resuming exact snapshots at any point during the run.

\subsection{Use of AI tools} \label{sec:ai}
AI tools were used to derive part of the solution and to write some of the code (Anthropic Claude Opus 5.5 and Fable 5.1; OpenAI Astra 6.0), and to spell-check the manuscript (Google Gemini Argon 4.0). The author checked all results and takes full responsibility for the content of this paper.

%% ======================================================================
\section{Accuracy} \label{sec:accuracy}
%% ======================================================================

\subsection{Trajectory fidelity and secular frequencies} \label{sec:divergence}\label{sec:yardstick}\label{sec:unbiased}
\begin{figure*}
\centering
\includegraphics[width=\textwidth]{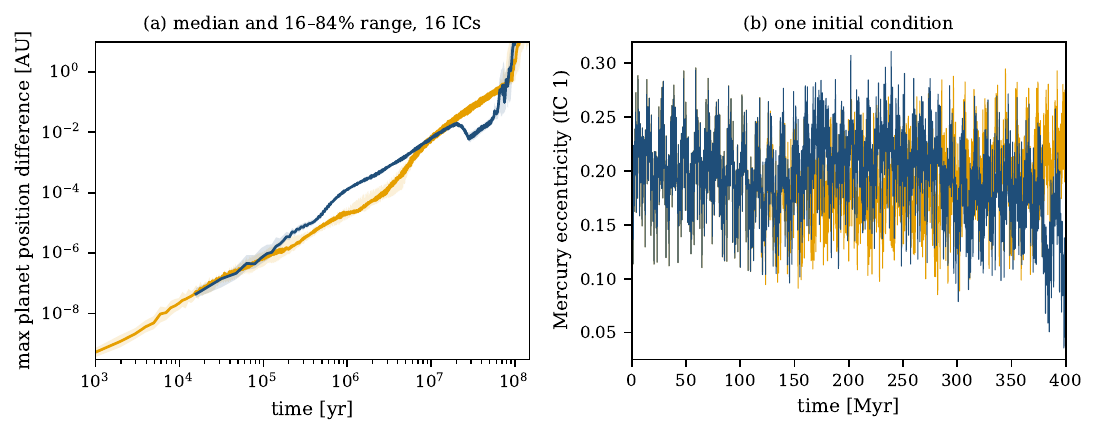}
\caption{(a) Largest planet position difference between pairs of trajectories from identical initial conditions (median and 16--84\% range of 16 pairs): two serial integrations that differ only in floating-point round-off (orange), and MEP against the serial integration (blue). (b) Mercury's eccentricity for one initial condition, serial (orange) and MEP (blue): the trajectories agree until chaos separates them, with the same character afterwards.}
\label{fig:divergence}
\end{figure*}
Because the Solar System is chaotic ($\lambda_{\rm L}\approx5$\,Myr), even two serial integrations that differ only in floating-point round-off diverge to $\sim$1\,AU within $\sim$100\,Myr (Figure~\ref{fig:divergence}a, orange). In MEP (blue), stopping each 16-slice window at $\tau=2\times10^{-6}$\,AU leaves a small residual below $\tau$: the parallel trajectory follows the round-off floor for the first $\sim$$2\times10^5$\,yr and remains within an order of magnitude of the serial round-off twin until both saturate near $100$\,Myr. Once separated by chaos, the trajectories exhibit the same qualitative orbital behaviour (Figure~\ref{fig:divergence}b).

Over $10$\,Myr intervals, before trajectories decorrelate, we measure the fundamental secular frequencies $g_i$ and $s_i$ via frequency analysis \citep[NAFF;][]{Laskar1993naff}. Compared against a reference serial integration at $\Delta t=3$\,d, the serial integrator at its standard $\Delta t=6$\,d step has a finite-step truncation error $\propto \Delta t^2$ of up to $5.1\times10^{-6}$\asec{} (in Mars' $g_4$). This truncation error---not floating-point round-off---is the physical yardstick for frequency accuracy. Across seven realizations of the initial conditions, MEP adds a random scatter of only $1.9\times10^{-8}$\asec{} rms to the inner planets' secular frequencies and a systematic shift of at most 2\% of the serial $\Delta t=6$\,d truncation error. Using larger windows ($N=64$ slices) reduces the frequency scatter by a further factor of three because most slices undergo additional iterations before the window terminates.

\subsection{Energy and angular-momentum conservation} \label{sec:energy}
\begin{figure*}
\centering
\includegraphics[width=\textwidth]{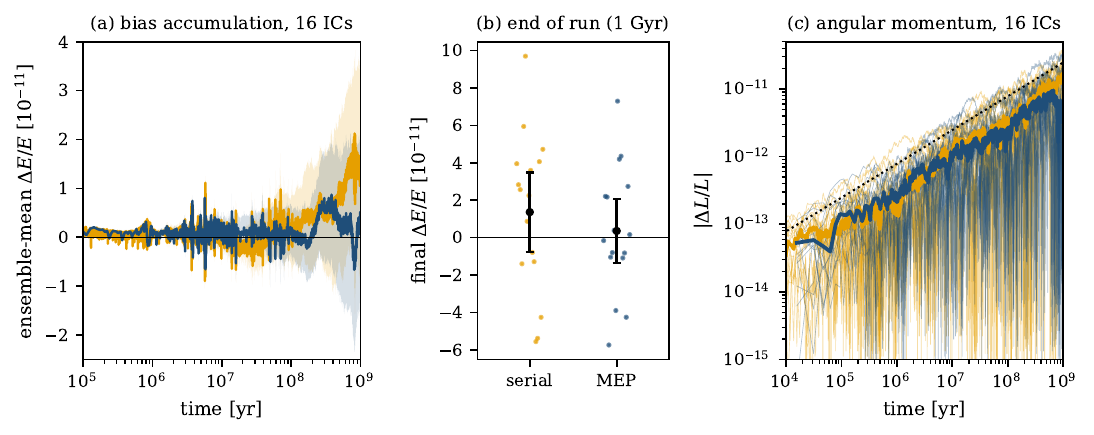}
\caption{Energy and angular-momentum conservation of 16-member ensembles over 1\,Gyr for the serial integrator (orange) and MEP (blue). (a) Ensemble mean of the relative energy error with $\pm2$ standard errors (shaded). (b) Final relative energy error of each run at 1\,Gyr, with ensemble mean and $\pm2$ standard errors. (c) Relative angular-momentum error of the 16 serial and 16 MEP runs (thin lines: individual runs; thick: median); the dotted line shows the Brouwer-law round-off growth ($10^{-16}\sqrt{N_{\rm steps}}$).}
\label{fig:bias}
\end{figure*}
To test conservation over billion-year timescales \citep[cf.][]{DagliRein2026}, we integrate a 16-member ensemble of the Solar System for $1$\,Gyr both serially and with MEP. After velocity projection at slice boundaries (Section~\ref{sec:projection}), boundary energy jumps remain at the floating-point round-off floor ($\sim$$5\times10^{-14}$ rms). Over the full $1$\,Gyr integrations, both the relative energy error and the relative angular-momentum error follow the unbiased $\propto 10^{-16}\sqrt{N_{\rm steps}}$ random walk of Brouwer's law \citep{Brouwer1937} (Figure~\ref{fig:bias}). At $1$\,Gyr, the ensemble-mean relative energy error is $(+0.4\pm0.9)\times10^{-11}$ for MEP against $(+1.4\pm1.1)\times10^{-11}$ for the serial ensemble, and the relative angular-momentum error is $(+0.1\pm0.4)\times10^{-11}$ against $(+0.3\pm0.5)\times10^{-11}$ serially. The paired differences between MEP and serial runs from the same initial conditions are consistent with zero within $1\sigma$.

\subsection{Long-term ensemble statistics} \label{sec:ensemble}
Because individual trajectories decorrelate after $\sim$$100$\,Myr, billion-year accuracy must be evaluated statistically across ensembles \citep{LaskarGastineau2009,Abbot2023,ReinBrownKanda2025}. While directly sampling the rare $\sim$1\% probability of a planetary collision over 5\,Gyr requires $\gtrsim10^3$ integrations, \citet{ReinBrownKanda2025} showed that long-term stability is governed by chaotic diffusion driven by the fundamental secular frequencies $g_i$ and $s_i$ and the resulting eccentricity excursions of Mercury. For the 16-member $1$\,Gyr ensembles above, two-sample Kolmogorov--Smirnov (KS) tests comparing the distributions of Mercury's maximum eccentricity $e_{\max}$ and the eight inner secular frequencies $g_1\ldots g_4$ and $s_1\ldots s_4$ against the serial reference ensemble show no statistically significant differences, with $p\ge0.21$ across all nine observables (Appendix~\ref{app:stats}, Figure~\ref{fig:stats}). Over $1$\,Gyr, Mercury's mean maximum eccentricity is $0.316\pm0.004$ for MEP versus $0.324\pm0.009$ for the serial ensemble ($p=1.00$). The parallel ensemble is statistically indistinguishable from the serial reference.

%% ======================================================================
\section{Performance} \label{sec:performance}
%% ======================================================================

\subsection{Wall-clock speed-up and convergence} \label{sec:speed}
On 16 cores, MEP converges in an average of $\Kbar=3.63$ iterations per window and achieves a measured wall-clock speed-up of 3.3 (3.29--3.32 across three repetitions of a 100-window, 1.58\,Myr benchmark) over a serial integration with the same C kernel, reducing a 5\,Gyr integration of the Solar System from 5.8\,h to 1.8\,h. On 64 cores, the integration completes in 52\,min, a speed-up of 6.7 over a single core of the C kernel. Compared to the 9.0\,h single-core record of \citet{DagliRein2026}, MEP reduces the wall-clock time at identical accuracy by a factor of 5.0 on 16 cores and 10.4 on 64 cores.

The number of iterations required per window is governed by the higher-order perturbations omitted from the first-order coarse model. At termination, the largest boundary update always occurs in the mean longitude of Venus, Earth, or Mars, where Keplerian shear (Equation~\ref{eq:shear}) amplifies the residual semi-major-axis error: after three iterations, a residual error of $2.9\times10^{-10}$ in Venus' $\ln a$ produces the measured phase update of $4.4\times10^{-6}$\,rad. The dominant omitted perturbation is the second-order Earth--Mars--Jupiter three-body near-resonance $4\lambda_{\rm E}-8\lambda_{\rm M}+3\lambda_{\rm J}$, which drives an oscillation in Mars' mean semi-major axis with a period of $\approx1836$\,yr and a relative amplitude of $2.2\times10^{-7}$---the largest long-period term in Mars' semi-major axis in the planetary theory VSOP87 (period 1783\,yr, amplitude $1.9\times10^{-7}$; \citealt{BretagnonFrancou1988}). Including second-order mass terms in the coarse propagator reduces the average iteration count to $\Kbar\approx2.4$, at the expense of a heavier serial coarse step (Section~\ref{sec:outlook}).

\subsection{Runtime breakdown and parallel overhead} \label{sec:budget}
\begin{figure}
\centering
\includegraphics[width=0.7\textwidth]{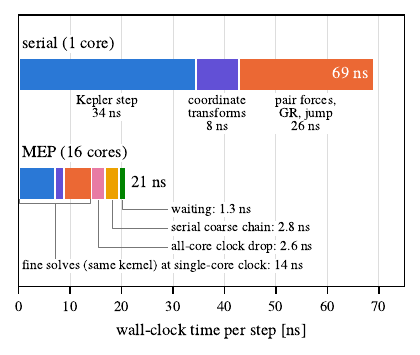}
\caption{Wall-clock time per integration step for the Solar System (eight planets with GR, $\Delta t=6$\,d), serial and with MEP on 16 cores (speed-up 3.3). Serial: 5.8\,h for 5\,Gyr with the C kernel, split by the cost fractions of its parts. The Kepler step comprises the series start, one evaluation of the Stiefel functions and the fifth-order correction; the interaction step comprises the transformations between Jacobi and heliocentric coordinates (prefix sums over the SIMD lanes) and the planet-pair forces with the GR terms and the jump step. MEP: one window (20\,ms for $16\times6\times10^4$ steps), with the fine solves split in the same proportions.}
\label{fig:steptime}
\end{figure}

Figure~\ref{fig:steptime} compares the effective wall-clock time per integration step on one core and on 16 cores. In a serial integration (69\,ns per step), the Kepler solver accounts for half of the runtime (34\,ns), pairwise force evaluations and GR corrections for 38\% (26\,ns), and Jacobi coordinate transformations for the remaining 12\% (8\,ns). On 16 cores, MEP completes a 16-slice window ($9.6\times10^5$ fine steps) in 20\,ms, reducing the effective wall-clock time per step to 21\,ns. Of this 21\,ns, the parallel fine integrations account for 16.8\,ns (80\% of the runtime): 14\,ns at the single-core clock rate plus 2.6\,ns from the processor's clock-frequency reduction under all-core AVX-512 load (5.3\,ms versus 4.4\,ms per fine slice). The remaining 4.1\,ns of parallel overhead comprises 2.8\,ns on the serial coarse chain and 1.3\,ns of thread synchronization and load-imbalance waiting across slices. Because the fine solver still dominates the wall-clock budget, any future acceleration of the fine kernel transfers almost directly to MEP.

\subsection{Parallel scaling} \label{sec:scaling}
\begin{figure}
\centering
\includegraphics[width=0.7\textwidth]{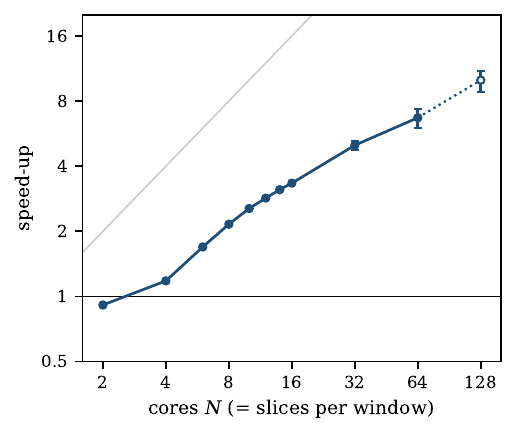}
\caption{Measured speed-up of MEP as a function of the number of cores $N$ (= slices per window) up to 64 cores, and the estimated speed-up for 128 cores (open symbol; bars: range of the estimate). The grey line marks ideal linear scaling.}
\label{fig:scaling}
\end{figure}

Figure~\ref{fig:scaling} shows the parallel scaling of MEP from $N=2$ to 64 cores, with one time slice assigned to each core. Two cores cannot provide a speed-up ($S=0.91$) because two parareal iterations already render both slices exact. Beyond two cores, the measured speed-up rises steadily to 3.3 on 16 cores, 5.0 on 32 cores, and 6.7 on 64 cores (52\,min for 5\,Gyr, or $10.4\times$ faster than the 9.0\,h record of \citealt{DagliRein2026}), with a gentle curvature caused by the gradual increase in the mean iteration count ($\Kbar$ rising from 3.0 at $N\le6$ to 3.6 at $N=16$) and by the growing length of the serial coarse chain. Up to 16 cores, the measured performance attains 85\%--91\% of the theoretical pipelined schedule (Equation~\ref{eq:events}) evaluated with ideal single-core slice costs; the difference arises primarily from all-core clock throttling, with a smaller contribution from helper-thread contention during coarse steps.

Extrapolating the schedule model to 128 cores---with $\Kbar$ increasing by 0.6--0.9 iterations per doubling of $N$---projects a speed-up of 10 (9--11) relative to a single core of the C kernel (open symbol in Figure~\ref{fig:scaling}).

%% ======================================================================
\section{Discussion} \label{sec:discussion}
%% ======================================================================

\subsection{Regime of applicability} \label{sec:use}
MEP reduces the wall-clock time of an individual trajectory rather than the total CPU cost of an ensemble: on 16 cores, it expends $4.8\times$ more core-hours than a single-core serial integration. For large ensembles whose members can all run concurrently, assigning one serial integration to each core remains the most economical strategy. Time parallelism becomes advantageous whenever a single long trajectory is the scientific product or the bottleneck: integrations spanning 10--100\,Gyr \citep{Hoang2022,DagliRein2026}, astrochronological orbital solutions, sequential workflows such as rare-event splitting \citep{Abbot2021} whose stages depend on one another, ensembles with fewer members than available cores, and rapid turnaround for exploratory runs. A 10\,Gyr integration that previously required ``several weeks'' on one core \citep{ReinTamayo2017sa}, or 5\,Gyr runs with ``a minimum wall-time of one month, even if a large number of cores is available'' \citep{BrownRein2020}, complete in 3.5\,h and 1.8\,h on 16 cores (and in 1.7\,h and 52\,min on 64 cores); a direct 100\,Gyr integration takes 17\,h on 64 cores. On a cluster, time parallelism and ensemble parallelism combine naturally, with one ensemble member assigned to each multi-core node and each trajectory accelerated across the cores of that node.

Because our speed-ups are measured against the fastest available serial algorithm for this problem, they do not stem from a slow baseline. Strictly speaking, MEP is not a symplectic map: like any parareal scheme, its accepted iterate equals the symplectic fine trajectory plus a residual bounded by $\tau$. However, as demonstrated in Section~\ref{sec:accuracy}, this residual preserves Brouwer-law conservation of energy and angular momentum over billion-year timescales and shifts the secular frequencies by at most 2\% of the serial integrator's own finite-step truncation error.

\subsection{Physical mechanism of convergence} \label{sec:insight}
In the parallel-in-time literature, parareal is commonly said to fail for oscillatory systems because the coarse propagator accumulates phase errors \citep{GanderHairer2014,Ruprecht2018}. For planetary orbits, this diagnosis is incomplete: state-independent phase offsets cancel out in the parareal update (Section~\ref{sec:parareal}), and convergence is governed instead by how accurately the coarse propagator resolves the sensitivity of the semi-major axes to the orbital state, which Keplerian shear amplifies by four orders of magnitude per kyr. Consequently, neither a higher-order coarse integrator (which would cost nearly as much as the fine solver) nor an artificially slowed orbital phase \citep[][which preserves secular evolution but amplifies short-period fluctuations]{MikkolaAarseth1996,WangNitadoriMakino2020} can resolve the bottleneck. Instead, the solution is a canonical transformation to mean elements that strips the short-period oscillations from the semi-major axes, drawing on two centuries of perturbation theory in celestial mechanics \citep{Brouwer1959,Deprit1969,Laskar1990,MurrayDermott1999}. The shear-amplified threshold of $\lesssim10^{-8}$ in the semi-major axes then dictates precisely which perturbative terms the coarse model must retain.

\subsection{Prospects for further acceleration and generalization} \label{sec:outlook}
Throughout this work, we have required MEP to match the full accuracy of the $\Delta t=6$\,d serial integrator ($\tau=2\times10^{-6}$\,AU, $K_{\min}=3$). However, \citet{ReinBrownKanda2025} recently demonstrated that Solar System ensembles remain statistically converged for serial time steps up to $\Delta t\approx32$\,d. Relaxing the parareal stopping criterion exploits this margin for an immediate 20\%--40\% additional speed-up while remaining well within the regime of statistical convergence. Specifically, loosening $\tau$ to $10^{-5}$--$3\times10^{-5}$\,AU at $K_{\min}=3$ lowers the mean iteration count to $\Kbar=3.0$--$3.2$ and increases the 16-core speed-up to $3.7$--$3.9$ ($1.5$\,h for 5\,Gyr), while keeping secular-frequency shifts below 11\% of the serial $\Delta t=6$\,d error. Allowing windows to terminate after $K_{\min}=2$ iterations at $\tau=3\times10^{-4}$\,AU (combined with a 180\,d giant-planet coarse step and an online linear least-squares calibration of the inner planets' secular rates across windows) lowers $\Kbar$ to 2.36 and yields a speed-up of 4.6 on 16 cores ($1.3$\,h for 5\,Gyr, or 39\,min on 64 cores), with frequency errors comparable to a $\Delta t\approx8$\,d serial run. Exploring optimal trade-offs between stopping tolerance and physical fidelity offers a straightforward path to further speed-ups.

Equation~(\ref{eq:speedup}) bounds the speed-up by $\min(N/K,1/r)$. Beyond the core-count scaling of Section~\ref{sec:scaling}, further gains are limited by the growth of $\Kbar$ and by the serial coarse chain ($N$ hops per window). Including second-order mass terms in the coarse propagator reduces $\Kbar$ to $\approx2.4$ even at strict tolerances and becomes net-beneficial once a coarse hop takes $\lesssim0.21$\,ms, which may be achievable by fusing passes over the pair tables. For still larger core counts, natural extensions include multilevel PinT schemes in which a fast secular model drives the mean-element coarse level in the spirit of PFASST and MGRIT \citep{EmmettMinion2012,Falgout2014}, or low-rank quasi-Newton updates to the coarse propagator along the inner planets' $(\ln a,\lambda)$ directions \citep{Vargas2023}, whose cost falls on the parallel fine solves rather than the serial coarse chain.

Finally, while our hybrid coarse propagator takes advantage of the architecture of the Solar System---four inner planets treated by perturbation theory and four well-separated giants integrated directly---the underlying grid-based perturbation theory is general and requires no power-series expansion in eccentricity or inclination. As a result, the coarse transformation remains accurate even during large secular eccentricity excursions of Mercury, provided orbits do not cross; if a trajectory enters an orbit-crossing close encounter where first-order perturbation theory no longer contracts the window within $K_{\max}$, the window either converges slice-by-slice via the exact fine propagator or can fall back to serial integration once an encounter is flagged. Extending MEP to compact exoplanetary systems near mean-motion resonances, such as TRAPPIST-1 or the resonant chains discovered by TESS and CHEOPS \citep{Tamayo2017trappist,Leleu2021}, will require integrating the resonant angles dynamically within the coarse propagator, just as we do here for the outer giants.

%% ======================================================================
\section{Conclusions} \label{sec:conclusions}
%% ======================================================================

We have demonstrated that a single long-term integration of the Solar System can be parallelized across time at the full accuracy of the serial integrator. The enabling physical insight is that parareal convergence for planetary orbits is governed by how accurately the coarse propagator resolves the semi-major axes, where Keplerian shear amplifies any discrepancy by four orders of magnitude per kyr. By transforming the inner planets to mean elements via a numerical first-order perturbation theory evaluated on the fly while integrating the outer giants directly at a large time step, our coarse propagator satisfies the resulting $\lesssim10^{-8}$ accuracy threshold at 2\% of the fine solver's cost and converges in 3--4 iterations.

Combined with a portable, latency-optimized C kernel for \whf{} that shortens the serial instruction dependency chain (reducing a single-core 5\,Gyr integration to 5.8\,h), mean-element parareal integrates the eight planets with general relativistic corrections for 5\,Gyr in 1.8\,h on 16 cores (a speed-up of 3.3 over one core of the C kernel, and $5.0\times$ faster than the 9.0\,h record of \citealt{DagliRein2026}) and in 52\,min on 64 cores ($6.7\times$ and $10.4\times$ faster, respectively). Across billion-year timescales, the parallel trajectories preserve Brouwer-law conservation of energy and angular momentum, reproduce the secular frequencies to within 2\% of the serial integrator's finite-step truncation error, and yield ensemble statistics indistinguishable from serial runs; relaxing the stopping tolerance within the regime of statistical convergence \citep{ReinBrownKanda2025} provides an additional 20\%--40\% speed-up. With single-core Wisdom--Holman integrators now near their hardware limits \citep{DagliRein2026}, time parallelism guided by celestial mechanics opens a new dimension for accelerating planetary dynamics.

\backmatter

\bmhead{Acknowledgements}
We thank Hanno Rein and the \rebound{} developers for making \rebound{} and \whf{} freely available. This work made use of \rebound{} \citep{ReinLiu2012}, \whf{} \citep{Javaheri2023,DagliRein2026}, Parachron (this work; \url{https://github.com/hippke/parachron}), NumPy, SciPy, Matplotlib.

\section*{Statements and Declarations}

\bmhead{Funding}
The author did not receive support from any organization for the submitted work.

\bmhead{Competing interests}
The author has no relevant financial or non-financial interests to disclose.

\bmhead{Ethics approval and consent to participate}
Not applicable.

\bmhead{Consent for publication}
Not applicable.

\bmhead{Data and code availability}
MEP is available in Parachron, a modified version of \rebound{}, at \url{https://github.com/hippke/parachron} (Python: \texttt{parachron.MEP}; C: \texttt{reb\_mep\_*}).

\bmhead{Author contribution}
Not applicable (single author).

%% CeMDA guidelines: appendix figures continue the numbering of the main text (no A1, A2, ...);
%% tables and equations follow the same convention
\edef\mainfigures{\the\value{figure}}\edef\maintables{\the\value{table}}\edef\mainequations{\the\value{equation}}
\begin{appendices}
\setcounter{figure}{\mainfigures}\renewcommand{\thefigure}{\arabic{figure}}
\setcounter{table}{\maintables}\renewcommand{\thetable}{\arabic{table}}
\setcounter{equation}{\mainequations}\renewcommand{\theequation}{\arabic{equation}}

\section{Hand-written assembly and compiled C} \label{app:asm}
\citet{DagliRein2026} rewrote \whf{} in x86 assembly to control register allocation and instruction scheduling by hand, arguing that compilers cannot match hand-written assembly for this kernel. Parachron's C kernel began as a direct port of their assembly kernel, permitting an apples-to-apples comparison of the same operations before our algorithmic modifications. This comparison shows no measurable advantage from hand-written assembly, for four reasons. First, for identical mathematical operations, the compiler generates equally fast code: the compiled Kepler step, which accounts for roughly two thirds of the runtime, required 114\,ns per step compared to 122\,ns for the hand-written assembly. Second, execution speed is governed by data-dependency chains rather than instruction ordering: the assembly kernel retires roughly one instruction per clock cycle on out-of-order processors capable of dispatching six or more, and the hardware scheduler reorders independent instructions dynamically at runtime. Third, hand-written assembly hardcodes microarchitectural trade-offs that are suboptimal on newer processors: for example, \citet{DagliRein2026} replace hardware square roots and divisions with an approximate reciprocal square root (\texttt{vrsqrt14pd}) followed by Newton--Raphson refinement, whereas on newer microarchitectures native square-root and division instructions make the full step $\sim$10\% faster---a compile-time switch in C, but a rewrite in assembly. Fourth, manual bit-level tricks can compromise numerical edge cases for negligible gain: the assembly kernel halves floating-point numbers by subtracting one from their binary exponent, which saves less than 1\% of the step time but maps $0.0$ to $-\infty$. Finally, whereas x86 assembly is restricted to AVX-512 processors, a single C implementation compiles cleanly across AVX-512, AVX2, and ARM architectures with bitwise identical results.

\begin{figure*}
\centering
\includegraphics[width=\textwidth]{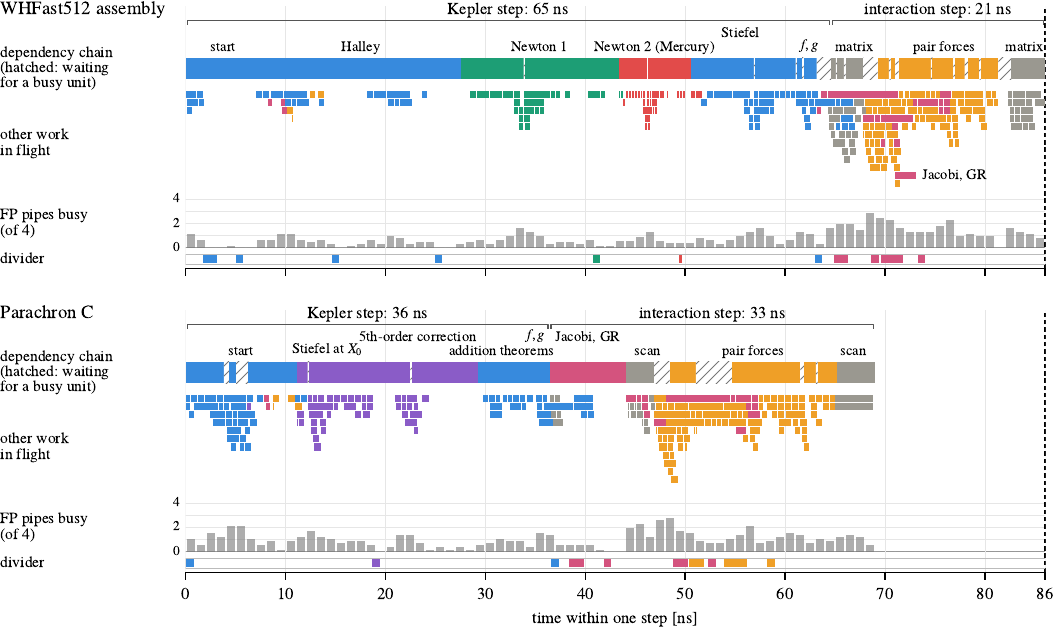}
\caption{Instruction-level schedule of one serial step for the Solar System (eight planets with GR, $\Delta t=6$\,d) for the assembly kernel of \citet{DagliRein2026} (top, 86\,ns average step, with Mercury's second Newton iteration weighted by its 46\% occurrence rate) and Parachron's C kernel (bottom, 69\,ns). Dependency chain: the critical path of sequentially dependent instructions that determines the step time; each bar represents the latency of one instruction, hatched where the chain stalls waiting for a busy functional unit. Other work in flight: non-critical instructions executing concurrently. FP pipes busy: average number of active floating-point pipes (out of four) per ns. Divider: occupancy of the non-pipelined division/square-root unit. Schedule modelled on an ideal out-of-order core (four 512-bit floating-point pipes, six instructions dispatched per cycle) executing the traced instruction stream of each kernel.}
\label{fig:schedule}
\end{figure*}

Because \whf{} processes the eight planets in parallel across the lanes of a single SIMD register, its step time is limited by instruction latency along the critical dependency chain---a few clock cycles per multiplication or addition and 10--20 cycles per division or square root---while the processor's remaining arithmetic units sit idle \citep[][Figure~\ref{fig:schedule}]{HennessyPatterson2019}. In Parachron's C kernel, we shorten this dependency chain in three ways. First, in the universal-variable Kepler solver \citep{StiefelScheifele1971}, we replace the two Halley and one to two Newton iterations of \whf{} with a sixth-order series reversion of Kepler's equation in $\Delta t$ for the starting value \citep[cf.][]{Danby1987} followed by a single fifth-order correction generalizing \citet{DanbyBurkardt1983}; all required derivatives follow from a single evaluation of the Stiefel functions at the starting value, and the Stiefel functions at the corrected value are obtained via their addition theorems rather than a second series evaluation. The Stiefel power series themselves are evaluated with Estrin's scheme \citep{Estrin1960,Knuth1997}, which reduces an $n$-term polynomial from $n$ sequential steps to $\approx\log_2 n$ parallel stages. Second, in the pairwise force evaluation, inverse square roots are computed using hardware square root and division for half of the planet pairs and Newton--Raphson refinement for the other half, balancing the load between the divider and the fused multiply-add units. Third, the transformations between Jacobi and heliocentric coordinates are evaluated as three-stage SIMD prefix-sum scans across vector lanes \citep{HillisSteele1986} instead of $8\times8$ matrix multiplications. In a $10^6$-step Kepler-only test from 4096 random phases ($\Delta t=3$--8\,d), the semi-major-axis drift of the C kernel is consistent with zero at the $10^{-21}$ level per step, whereas the assembly kernel exhibits a drift of up to $4\times10^{-20}$ per step for Mercury.

Figure~\ref{fig:schedule} compares a single step of the assembly kernel of \citet{DagliRein2026} and Parachron's C kernel instruction by instruction; shortening the dependency chain reduces the Kepler step from 65\,ns to 36\,ns on average and the full step from 86\,ns to 69\,ns. Wall-clock timings up to 16 cores were measured on an AMD Ryzen~9 9950X (16 cores, Zen~5, full-width 512-bit AVX-512) and 32- and 64-core timings on an AMD EPYC~9575F (64 cores, Zen~5). Additional validation and portability tests were performed on an Intel Xeon Scalable processor (Sapphire Rapids, native AVX-512), an AMD Ryzen~9 7950X (Zen~4, which executes AVX-512 instructions via double-pumped 256-bit execution units), an AMD Ryzen~9 5950X (Zen~3, AVX2 only), hybrid Intel processors combining Performance (P) and Efficient (E) cores without AVX-512 (Intel Core~i5-13500 and Core Ultra~7 255), and an Apple~M1 processor (ARM64 NEON).

\section{Synchronization energy offset in \whf{}} \label{app:syncbias}
\begin{figure}
\centering
\includegraphics[width=0.63\textwidth]{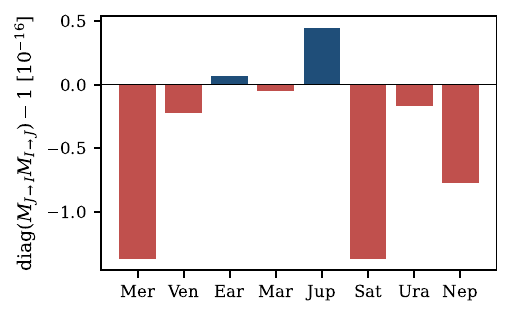}
\caption{Deviation from unity of the diagonal entries of the product of \whf{}'s precomputed Jacobi-to-inertial and inertial-to-Jacobi transformation matrices for the Solar System masses. Each synchronization rescales the Jacobi coordinates of each planet by these factors, adding $+7.5\times10^{-17}$ to the relative energy per call.}
\label{fig:syncbias}
\end{figure}

Because parareal restarts the fine integrator at every slice boundary, MEP exposed a small systematic energy bias in upstream \whf{} that also affects serial integrations with frequent output. Every synchronization call changes the relative energy of the Solar System by a constant offset $\Delta E/E=(+7.54\pm0.49)\times10^{-17}$ ($15\sigma$, reproduced with unmodified \rebound{}), independent of the orbital state. This offset arises from the pair of precomputed $8\times8$ matrices used to convert between inertial and Jacobi coordinates, which are not exact floating-point inverses: their product is diagonal to $10^{-19}$, but the diagonal entries differ from unity by $(-1.37,-0.22,+0.07,-0.05,+0.45,-1.37,-0.17,-0.77)\times10^{-16}$ for Mercury through Neptune (Figure~\ref{fig:syncbias}). Each round-trip coordinate transformation rescales planet $i$'s Jacobi position and velocity by $(1+\epsilon_i)$ and shifts its energy by $\approx2\epsilon_i|U_i|$; summing over the planets' potential energies predicts $+7.6\times10^{-17}$ per call, matching the measured offset. In a 5\,Gyr integration with output every kyr, this bias accumulates to $\Delta E/E\approx+4\times10^{-10}$, exceeding the Brouwer-law round-off random walk of $\sim$$10^{-16}N_{\rm steps}^{1/2}\approx5\times10^{-11}$ \citep{Brouwer1937,DagliRein2026}. Replacing the matrix multiplications at synchronization boundaries with the recursive Jacobi transformations of \rebound{} eliminates the offset ($\le1.5\times10^{-17}$ per call at $2\sigma$) at no measurable runtime cost.

\section{Pipelined schedule and parameter choices} \label{app:speedup}\label{sec:params}
With $t(k,n)$ denoting the wall-clock time at which the boundary state $U^k_{n+1}$ becomes available, the wave-pipelined schedule (Section~\ref{sec:pipeline}) obeys
\begin{equation}
t(k,n+1) = \max\!\left[t(k,n),\; t(k-1,n)+T_F(n)\right] + T_{\rm hop}.
\label{eq:events}
\end{equation}
In a window of $N$ slices, the initial coarse sweep takes $NT_G$. In iteration $k\ge1$, the coarse hop that produces $U^k_{n+1}$ cannot begin until the fine integration $F(U^{k-1}_n)$---which started when $U^{k-1}_n$ became available---has finished. When all fine solves have equal duration $T_F$, the wave of iteration $k$ trails that of iteration $k-1$ by $T_F+T_G$, so the final boundary of the window is ready after $NT_G+K(T_F+T_G)$, whereas the serial runtime for the window is $NT_F$; dividing the two yields Equation~(\ref{eq:speedup}). For $K\ll N$ and $r=T_G/T_F\ll1$, the speed-up approaches $N/K$; as $K\to N$ (no convergence), the speed-up falls below unity, as observed for a coarse WH integrator. When fine slice durations vary slightly, $T_F$ is governed by the slowest slice in each iteration.

Table~\ref{tab:params} lists the algorithmic parameters of MEP, the range explored for each, and the resulting performance. Each parameter was tuned using both schedule-model predictions from measured iteration counts and full benchmark integrations. Most parameters lie on a broad plateau where neighbouring values alter the wall-clock speed-up by only a few per cent. The default stopping tolerance ($\tau=2\times10^{-6}$\,AU) is chosen for physical accuracy rather than maximum speed: it is the loosest threshold whose added secular-frequency shift is negligible compared to the serial integrator's own finite-step truncation error (Section~\ref{sec:unbiased}).

\begin{table*}
\centering
\caption{Parameters of MEP: default, range explored, and outcome.}
\label{tab:params}
\footnotesize
\begin{tabular}{>{\raggedright\arraybackslash}p{2.4cm}>{\raggedright\arraybackslash}p{2.0cm}>{\raggedright\arraybackslash}p{2.4cm}>{\raggedright\arraybackslash}p{3.6cm}}
\hline
Parameter & Default & Explored & Outcome \\
\hline
slices per window $N$ & 16 (= cores) & 12--24; 32 on SMT threads & fewer slices with more helpers: 2.7--2.8; 20--24 slices: 2.8--2.9; 32 SMT threads: 2.2--2.3 \\
slice length $T$ & 986\,yr ($6\times10^4$ steps) & 246--9856\,yr & flat optimum: 789 and 1183\,yr 3--8\% slower, 1380\,yr 10\%; 246--493\,yr raise $r$, $\ge$10\,kyr raise $K$ \\
tolerance $\tau$ & $2\times10^{-6}$\,AU & $10^{-6}$--$3\times10^{-5}$\,AU & $\Kbar$ decreases from 3.63 at $2\times10^{-6}$\,AU to its floor of 3.0 at $3\times10^{-5}$\,AU (Section~\ref{sec:outlook}) \\
$K_{\min}$, $K_{\max}$ & 3, 6 & $K_{\min}=2$, 3 & $K_{\min}=2$ lowers $\Kbar$ to 2.4 when combined with a 180\,d giants' step and online secular calibration (Section~\ref{sec:outlook}) \\
stopping rule & largest update $\le\tau$ & error predicted from the contraction & no better accepted states at equal $\Kbar$ (625 windows) \\
Fourier grids (inner--inner, inner--giant) & $32^2$, $16^2$; Venus--Earth $16\times64$ synodic & $8^2$--$64^2$; synodic $8\times64$--$32\times64$ & smaller grids raise $K$; larger square grids lower $K$ but cost more on the coarse chain; the synodic Venus--Earth grid reaches the accuracy of $64^2$ for that pair at the cost of $32^2$ \\
giants' coarse step & 360\,d & 180--1000\,d & 240--450\,d within the scatter; $\ge$540\,d no convergence \\
coarse helper threads & 8, on SMT siblings of the first cores & 0--8; placements & pinned placement +5\% \\
\hline
\end{tabular}
\end{table*}

\section{Linearized error recursion and conservation projection} \label{app:jacobian}
Linearizing the parareal update (Equation~\ref{eq:parareal}) around the fine trajectory yields the error recursion
\begin{equation}
e^{k+1}_{n+1} = G'_n\, e^{k+1}_n + \left(F'_n - G'_n\right) e^k_n ,
\label{eq:linear}
\end{equation}
where $F'_n$ and $G'_n$ are the $48\times48$ Jacobians of the fine and coarse slice maps in Jacobi orbital elements. Convergence is therefore governed by the Jacobian mismatch $F'-G'$ rather than by state-independent errors in $G$. Evaluated with measured slice Jacobians, Equation~(\ref{eq:linear}) reproduces the nonlinear parareal iteration step by step. In osculating elements, short-period fluctuations produce off-diagonal fine-Jacobian entries $\partial\ln a_{\rm end}/\partial\lambda_j\approx10^{-6}$--$10^{-5}$ and, combined with Keplerian shear, $\partial\ln a_{\rm end}/\partial\ln a_j$ up to 0.09 (Earth with respect to Venus); in mean elements, the $\ln a$ rows of the slice Jacobian are close to the identity.

To project each corrected boundary state onto the target energy $E_{\rm target}=E(U^k_n)+[E(F(U^{k-1}_n))-E(U^{k-1}_n)]$ and target angular momentum $\mathbf L_{\rm target}=\mathbf L(U^k_n)+[\mathbf L(F(U^{k-1}_n))-\mathbf L(U^{k-1}_n)]$ (Section~\ref{sec:projection}), we adjust the barycentric velocities by $\delta\mathbf v_i=\alpha\,\mathbf v_i+\boldsymbol\beta\times\mathbf r_i$. The scalar $\alpha$ and three-vector $\boldsymbol\beta$ are obtained from two Newton iterations of the $4\times4$ linear system
\begin{equation}
\begin{pmatrix} 2E_{\rm kin} & \mathbf L^{\rm T}\\ \mathbf L & \mathsf I\end{pmatrix}\begin{pmatrix}\alpha\\ \boldsymbol\beta\end{pmatrix}=\begin{pmatrix}E_{\rm target}-E\\ \mathbf L_{\rm target}-\mathbf L\end{pmatrix},\qquad \mathsf I=\sum_i m_i\left(r_i^2\,\mathsf 1-\mathbf r_i\mathbf r_i^{\rm T}\right).
\end{equation}
Two numerical details matter at the $10^{-16}$ level. First, the velocity correction must be applied incrementally as $\mathbf v_i\leftarrow\mathbf v_i+\delta\mathbf v_i$: the mathematically equivalent multiplicative rescaling by $\sqrt{1+x}$ with $x\sim10^{-12}$ rounds asymmetrically near unity and removes $1.0\times10^{-16}$ of the energy per projection, causing a measurable secular drift. Second, angular momentum must be projected alongside energy; with energy projection alone, $|\mathbf L|$ drifts by $-3.1\times10^{-10}$\,Gyr$^{-1}$.

\section{Mean-element transformation and Gauss' equations} \label{app:gauss}
We represent each planet by its Jacobi equinoctial elements $E=(\ln a,\lambda,k,h,q,p)$, with $k+\mathrm{i}h=e\,\mathrm{e}^{\mathrm{i}\varpi}$, $q+\mathrm{i}p=\sin(i/2)\,\mathrm{e}^{\mathrm{i}\Omega}$, and mean longitude $\lambda$. For every pair of planets $i<j$, we evaluate the first-order perturbing accelerations in Jacobi coordinates on a regular grid of the two mean longitudes $(\lambda_i,\lambda_j)$, holding the slow elements ($\ln a,k,h,q,p$) fixed at their current values. For a planet with Jacobi position $\mathbf r$, velocity $\mathbf v$, and Keplerian parameter $\mu$ (the sum of the masses interior to and including the planet), the energy equation gives $\dd\ln a/\dd t=2a\,\mathbf v\cdot\mathbf a_{\rm pert}/\mu$ exactly. The rates of the remaining elements follow from Gauss' equations, $\dd E/\dd t=(\partial E/\partial\mathbf v)\cdot\mathbf a_{\rm pert}$, where the partial derivatives at fixed position are evaluated using forward-mode dual numbers through the Cartesian-to-element conversion (finite differences are accurate to only $\sim$$10^{-6}$ in some components). The interior planet of a pair experiences the direct term and the dipole indirect term, while the exterior planet experiences the interior mass distribution in the heliocentric-like Jacobi form; compared with a direct $N$-body sum, these accelerations agree to $3$--$5\times10^{-5}$ (relative) for the inner planets, the residual being second order in the masses.

A two-dimensional FFT of each element rate yields Fourier harmonics $f_{\mathbf k}$ with wave vector $\mathbf k=(k_1,k_2)$ and frequency $\nu_{\mathbf k}=k_1n_i+k_2n_j$, where $n_i$ and $n_j$ are the mean motions corresponding to the mean semi-major axes (evaluated iteratively from the same Fourier tables). For the close Venus--Earth pair, the grid is instead spanned by $\lambda_{\rm V}$ and the synodic angle $\lambda_{\rm V}-\lambda_{\rm E}$ with $16\times64$ points; its harmonic $(m,l)$ has wave vector $\mathbf k=(m+l,-l)$ and frequency $mn_{\rm V}+l(n_{\rm V}-n_{\rm E})$. Because the amplitude of a harmonic scales as $e^{|k_1+k_2|}$ \citep{MurrayDermott1999} but decays slowly with synodic order for a close pair, this skewed grid resolves harmonics $|m|\le8$, $|l|\le32$ at the same cost as a $32\times32$ square grid. The $\mathbf k=0$ mode gives the first-order secular rate, while the short-period displacement of each element at the current mean longitudes $\boldsymbol\theta=(\lambda_i,\lambda_j)$ is
\begin{equation}
\Delta E = \sum_{\mathbf k\neq0}\frac{f_{\mathbf k}\,\mathrm{e}^{\mathrm{i}\mathbf k\cdot\boldsymbol\theta}}{\mathrm{i}\nu_{\mathbf k}},
\qquad
\Delta\lambda_{\rm shear} = -\frac32\,n\sum_{\mathbf k\neq0}\frac{f^{(\ln a)}_{\mathbf k}\,\mathrm{e}^{\mathrm{i}\mathbf k\cdot\boldsymbol\theta}}{(\mathrm{i}\nu_{\mathbf k})^2},
\label{eq:disp}
\end{equation}
where $\Delta\lambda_{\rm shear}$ accounts for the perturbation of the Keplerian mean motion $n(a)$ by the short-period oscillation in $\ln a$. The GR correction enters as a one-planet table. Mean elements are obtained by subtracting the summed displacements from the osculating elements. In practice, the displacements of all elements are evaluated with a single FFT per planet pair, because $\sum_{\mathbf k}f_{\mathbf k}\mathrm{e}^{\mathrm{i}\mathbf k\cdot\boldsymbol\theta}/\nu_{\mathbf k}$ reduces to a grid sum of the element rates weighted by the inverse-transformed kernel $\mathrm{e}^{\mathrm{i}\mathbf k\cdot\boldsymbol\theta}/\nu_{\mathbf k}$.

\section{Ensemble statistical tests} \label{app:stats}
\begin{figure*}
\centering
\includegraphics[width=\textwidth]{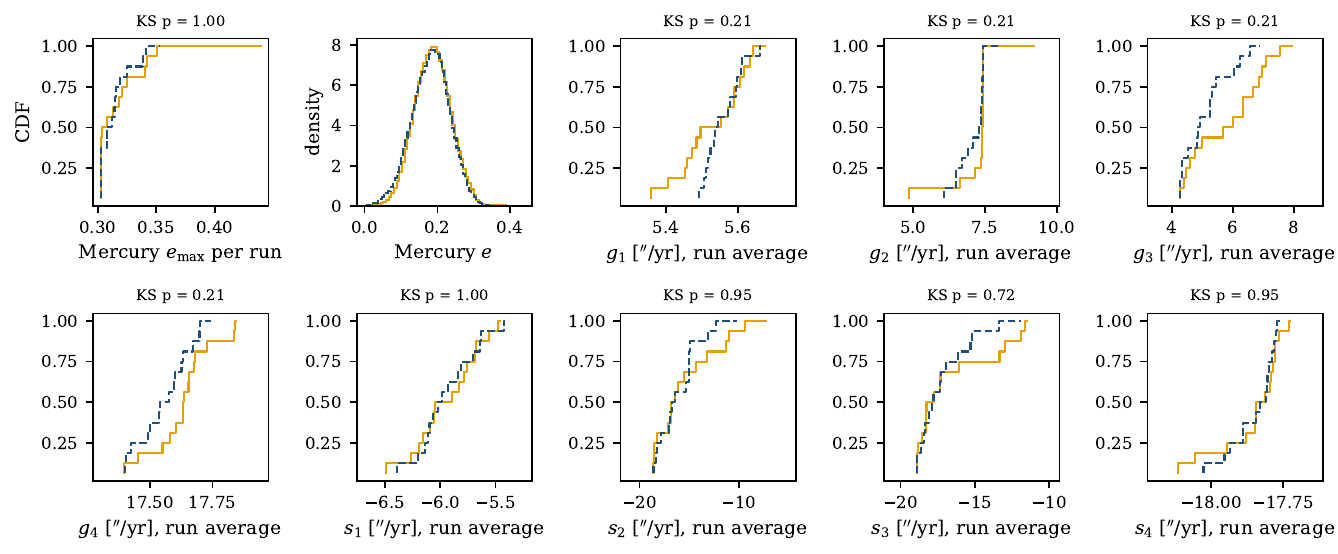}
\caption{Ensemble statistics over 1\,Gyr for the 16-member serial ensemble (orange) and MEP ensemble (blue). Top left: cumulative distribution of Mercury's maximum eccentricity per run; next: distribution of Mercury's eccentricity; remaining panels: cumulative distributions of the run averages of the inner secular frequencies $g_1\ldots g_4$ and $s_1\ldots s_4$, with two-sample KS $p$-values shown above each panel.}
\label{fig:stats}
\end{figure*}

Figure~\ref{fig:stats} shows the cumulative distributions of Mercury's maximum eccentricity $e_{\max}$ (after $50$\,Myr), Mercury's eccentricity distribution, and the run-averaged inner secular frequencies $g_1\ldots g_4$ and $s_1\ldots s_4$ (measured in $20$\,Myr windows) for the 16-member $1$\,Gyr serial and MEP ensembles of Section~\ref{sec:ensemble}. The two-sample KS $p$-values comparing the MEP ensemble against the serial ensemble are $1.00$ for $e_{\max}$ and span $0.21$--$1.00$ across all eight secular frequencies ($g_1\ldots g_4, s_1\ldots s_4$). None of the nine observables shows a statistically significant difference ($p<0.05$).

\end{appendices}

\bibliography{refs}

\end{document}